\documentclass[showpacs,aps,graphicx,twocolumn,]{revtex4}
\usepackage{mathrsfs}
\usepackage{amsfonts}
\usepackage{amssymb}
\usepackage{graphicx}
\usepackage{hyperref}
\usepackage{eufrak}
\usepackage{multirow}

\begin{document}

\title{Deterministic CNOT Gate on electron qubits using  quantum-dot spins in  double-sided optical microcavities}

\author{Hai-Rui Wei  and Fu-Guo Deng\footnote{Corresponding author: fgdeng@bnu.edu.cn}}
\address{ Department of Physics, Applied Optics Beijing Area Major
Laboratory, Beijing Normal University, Beijing 100875, China}

\date{\today }

\begin{abstract}
We propose a scheme to construct a deterministic CNOT gate on static
electron-spin qubits, allowing for deterministic scalable quantum
computing in solid-state systems.The excess electron confined in a
charged quantum dot inside a double-sided optical microcavity
represents the qubit, and the single photons play a medium role.
Moreover, our device can work in both the weak coupling and the
strong coupling regimes, but high fidelities are achieved only when
the ratio of the side leakage to the cavity loss is low. Finally, we
assess the feasibility of this device and show it can be implemented
with current technology.
\end{abstract}


\maketitle

The controlled-not (CNOT) gate has numerous applications in the
field of quantum information science and it is one of the elementary
elements for a quantum computer \cite{comput1,comput2,comput3}. In
1995, Barenco \emph{et al.} \cite{univer} proven that any $n$-qubit
quantum computation can be achieved by using a sequence of one-qubit
gates and CNOT gates. The archetypal two-qubit CNOT gate, or its
equivalents, have been demonstrated from various perspectives and
for different physical systems, including trapped ions
\cite{ion1,ion2}, nuclear magnetic spins \cite{NMR}, superconducting
circuits \cite{super1,super2}, and linear optics
\cite{linear1,linear2}. In fact, each of these systems has its
bottleneck. For example, based on  linear optical elements, the
maximum probability for achieving a CNOT gate is 3/4 \cite{KLM}. A
Superconducting circuit is fragile to decoherence. In 2004,
Beenakker \emph{et al.}  \cite{CNOT1} proposed a theoretic protocol
for CNOT gate on moving electrons.  Nemoto  and Munro  \cite{CNOT2}
introduced a protocol for a CNOT gate on photons with cross-Kerr
nonlinearity. The CNOT gate on static  qubits is more useful for a
scalable quantum computing.

Recent works show that the electron spin in a quantum dot (QD)
\cite{QD} can be used to store and process quantum information due
to the long electron-spin coherence time ($\sim$ $\mu$s)\cite{coher
time} using spin echo techniques, which is limited by the spin
relaxation time ($\sim$ ms) \cite{spin time}, and it hold great
promising in quantum communications, quantum information processing,
and quantum networks. The spin-QD-cavity unit, e.g., an electron
confined in a self-assembled In(Ga)As QD or a GaAs interface QD
inside a single-sided or a double-sided optical resonant cavity was
proposed by Hu \emph{et al.} \cite{Hu1,Hu2}. In this unit, the spin
represents the qubit and promises scalable quantum information
computing. A single spin qubit can be read out by the information of
a coupling photon, and spin manipulation is well developed using
pulsed magnetic-resonance technique. This unit has been used for
constructing a hybrid CNOT gate and a phase-shift gate, two-photon
 Bell-state analyzer (BSA), teleportation, entanglement
swapping, entanglement purification, and creating photon-photon,
photon-spin, and spin-spin entanglements
\cite{Hu1,Hu2,Hu3,Hu4,Appli1,Appli2,Appli3}.

In this paper,  we investigate the construction of a CNOT gate on
the two static electrons confined in two charged QDs inside two
double-sided microcavities. We first propose a device which can
convert the spin parity into the out-coming photon polarization
information. Using two such parity measurements, we construct a CNOT
gate on two static electron-spin qubits, resorting to an ancillary
static electron-spin qubit, a single-qubit measurement, and the
application of single-qubit operations. Moreover, a complete
deterministic two-spin BSA was constructed. In our scheme, the CNOT
gate promises a scalable quantum computing in solid-state systems,
in which two single photons only are mediums. The device works in
both the weak coupling and the strong coupling regimes, but high
fidelities are achieved only when the side leakage and cavity loss
is low.

\begin{figure}[!h]
\begin{center}
\includegraphics[width=4.5 cm,angle=0]{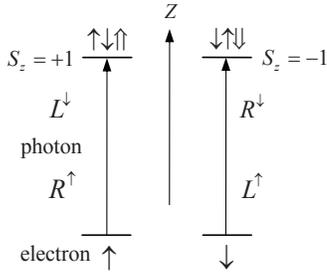}
\caption{Relevant energy levels and spin selection rules for optical
transition of negatively charged exciton  $X^-$.} \label{Fig1}
\end{center}
\end{figure}

The spin-QD-double-side-cavity unit, we consider here, is a singly
electron charged self-assembled GaAs/InAs interface QD inside an
optical resonant double-sided microcavity with two  partially
reflective mirrors. The potential of this system has also been
recognized in Ref.\cite{Hu2}. An exciton ($X^-$) that consists of
two electrons and a hole can be created by optical excitation. Here,
the dipole is resonant with  cavity mode, probed with a resonant
light. The four relevant electronic levels are shown in
Fig.\ref{Fig1}. \cite{Hu2}. Due to Pauli's exclusion principle,
there are two dipole transitions, one involving a photon with the
spin $s_z=+1$ and the other involving a photon with $s_z=-1$.
Considering  a photon with $s_z=\pm1$, if the injecting photon
coupled to the dipole, the cavity is reflected,  and both the
polarization and the propagation direction of the photon will be
flipped. Otherwise, the cavity is transmissive and the photon will
acquire a $\pi$ mod $2\pi$ phase shift relative to a reflected
photon. The rules of the input states changed under the interaction
of the photons with $s_z = \pm1$ and the cavity are described as
follows:
\begin{eqnarray}   \label{eq.1}
|R^\uparrow\uparrow\rangle&\rightarrow&|L^\downarrow\uparrow\rangle,\;\;\;\;\;\;\;\;\;
|L^\uparrow\uparrow\rangle\rightarrow-|L^\uparrow\uparrow\rangle, \nonumber\\
|R^\downarrow\uparrow\rangle&\rightarrow&-|R^\downarrow\uparrow\rangle,\;\;\;\;\;\;
|L^\downarrow\uparrow\rangle\rightarrow|R^\uparrow\uparrow\rangle,\nonumber\\
|R^\uparrow\downarrow\rangle&\rightarrow&-|R^\uparrow\uparrow\rangle,\;\;\;\;\;\;
|L^\uparrow\downarrow\rangle\rightarrow|R^\downarrow\downarrow\rangle,\nonumber\\
|R^\downarrow\downarrow\rangle&\rightarrow&|L^\uparrow\uparrow\rangle,\;\;\;\;\;\;\;\;\;
|L^\downarrow\downarrow\rangle\rightarrow-|L^\downarrow\downarrow\rangle.
\end{eqnarray}
Here, $|\uparrow\rangle$ and $|\downarrow\rangle$ represent the
electron-spin states $|+\frac{1}{2}\rangle$ and
$|-\frac{1}{2}\rangle$, respectively. The spin quantization axis for
angular momentum is along the normal direction of cavity that is the
$z$ axis. $|R\rangle$ ($|L\rangle$) is the right (left) circular
polarization of a photon, and the superscripts $\uparrow$ and
$\downarrow$ indicate the propagation directions of a photon along
the $z$ axis. In Fig.\ref{Fig1}, $|\Uparrow\rangle$ and
$|\Downarrow\rangle$ represent hole-spin states
$|+\frac{3}{2}\rangle$ and $|-\frac{3}{2}\rangle$, respectively.

\begin{figure}[!h]
\begin{center}
\includegraphics[width=8.0 cm,angle=0]{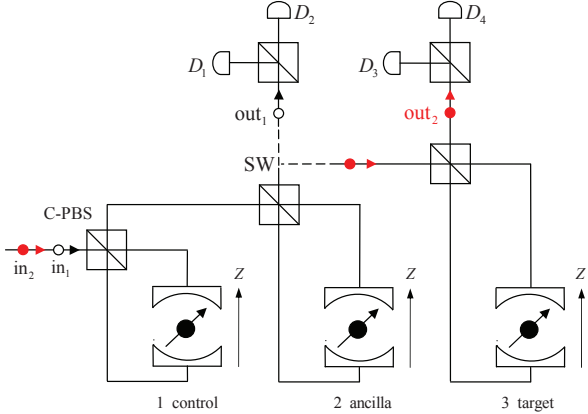}
\caption{(Color online) Scheme for  a CNOT gate on noninteracting
static electron-spin qubits inside the double-sided microcavities.
$D_i$ ($i=1,2,3,4$) are four single-photon detectors. The polarizing
beam splitter in the circular basis (C-PBS) transmits a
right-circular polarization photon $|R\rangle$  and reflects a
left-circular polarization photon $|L\rangle$. SW is an optical
switch. After the first PCG on the control qubit and the ancilla
qubit, the first C-PBS is rotated by $90^\circ$, so that the second
probe photon $|L\rangle$  (label $in_2$) deserts the first cavity
and injects directly into the cavities  2 and  3 in sequence.}
\label{Fig2}
\end{center}
\end{figure}

Now, let us describe the procedure for the construction of the
parity-check gate (PCG) and a CNOT gate for
spin-QD-double-side-cavity units.  Based on the rules discussed
above, the principle of our PCG for two spin qubits (in the first
two cavities) is shown in Fig.\ref{Fig2}. It is relied  on the
spin-to-polarization conversion.  Two excess electron spins in the
cavities are in two arbitrary states. A probe photon passes through
the polarizing beam splitter in the circuit basis (C-PBS) and it
injects into the first and the second cavities in succession. After
it interacts with the cavities, the photon is detected. By detecting
the output of the photon, one can distinguish the spin states of the
two-electron system \{$|\uparrow_1\uparrow_2\rangle,
|\downarrow_1\downarrow_2\rangle$\} from
\{$|\uparrow_1\downarrow_2\rangle,|\downarrow_1\uparrow_2\rangle$\}.
If the two spins are parallel ($|\uparrow_1\uparrow_2\rangle$ or
$|\downarrow_1\downarrow_2\rangle$), the polarization of the probe
photon  in the state $|R\rangle$ ($|L\rangle$) will  remain and the
photon will trigger the detector $D_2$ ($D_1$); otherwise, the state
of the probe photon will be flipped and the photon will be detected
by the detector $D_1$ ($D_2$). The evolution of the photon-cavity
state can be described as
\begin{eqnarray}  \label{eq.2}
|R^\downarrow\rangle|\uparrow_1\uparrow_2\rangle&\rightarrow&|R^\downarrow\rangle|\uparrow_1\uparrow_2\rangle,
\;\;|R^\downarrow\rangle|\uparrow_1\downarrow_2\rangle\rightarrow-|L^\uparrow\rangle|\uparrow_1\downarrow_2\rangle,\nonumber\\
|R^\downarrow\rangle|\downarrow_1\downarrow_2\rangle&\rightarrow&|R^\downarrow\rangle|\downarrow_1\downarrow_2\rangle,
\;\;|R^\downarrow\rangle|\downarrow_1\uparrow_2\rangle\rightarrow-|L^\uparrow\rangle|\downarrow_1\uparrow_2\rangle,\nonumber\\
|L^\uparrow\rangle|\uparrow_1\uparrow_2\rangle&\rightarrow&|L^\uparrow\rangle|\uparrow_1\uparrow_2\rangle,
\;\; |L^\uparrow\rangle|\uparrow_1\downarrow_2\rangle\rightarrow-|R^\downarrow\rangle|\uparrow_1\downarrow_2\rangle,\nonumber\\
|L^\uparrow\rangle|\downarrow_1\downarrow_2\rangle&\rightarrow&|L^\uparrow\rangle|\downarrow_1\downarrow_2\rangle,
\;\;|L^\uparrow\rangle|\downarrow_1\uparrow_2\rangle\rightarrow-|R^\downarrow\rangle|\downarrow_1\uparrow_2\rangle.\nonumber\\
\end{eqnarray}

Based on spin-QD-double-side-cavity systems, the principle of our
CNOT gate is shown in Fig.\ref{Fig2}. It is used to flips the spin
of the target qubit if the spin of the control qubit is
$|\downarrow\rangle$; otherwise, it does nothing.  Suppose that the
two excess electron  spins in the first cavity and third cavity are
considered as the  control qubit and the target qubit, respectively.
They are  in two  arbitrary states
$|\psi^s_1\rangle=\alpha_1|\uparrow_1\rangle+\beta_1|\downarrow_1\rangle$
and
$|\psi^s_3\rangle=\alpha_3|\uparrow_3\rangle+\beta_3|\downarrow_3\rangle$,
respectively. The ancilla qubit in the second cavity is prepared in
the state
$|\psi^s_{anci}\rangle=\frac{1}{\surd2}(|\uparrow_2\rangle+|\downarrow_2\rangle)$.
Our scheme consists of three parts. (i) We take two PCGs on spin
pairs 1-2 and 2-3 in series, with a Hadamard transformation (e.g.,
using a $\pi/2$ microwave pulse)
\begin{eqnarray}    \label{eq.3}
|\uparrow\rangle\rightarrow\frac{1}{\surd2}(|\uparrow\rangle+|\downarrow\rangle),\;\;\;
|\downarrow\rangle\rightarrow\frac{1}{\surd2}(|\uparrow\rangle-|\downarrow\rangle),
\end{eqnarray}
on the ancilla qubit and target qubit before and after the second
PCG operation, respectively. The first probe photon (label $in_1$)
is originally in state $|R^\downarrow_1\rangle$ and the second one
(label $in_2$) is in $|L^\uparrow_2\rangle$. (ii) The ancilla qubit
is measured. (iii) According to the result of two PCGs and the spin
of the ancilla qubit, a proper classical feed-forward is performed
on the control qubit and the target qubit to complete a CNOT gate
with the success probability of $100\%$. The correspondences between
the results of each measurements and specific feed-forwards are
given in Table \ref{Table}.

\begin{table}[!h]
 \centering
\newcommand{\minitab}[2][c]{\begin{tabular}{#1}#2\end{tabular}}
\caption{The correspondences between the results of two PCG
operations and the spin of the ancilla and the feed-forward
operators applied to the control and the target spins
 in the construction of a static two-spin-qubit CNOT gate.}
\def\temptablewidth{0.45\textwidth}
\begin{tabular}{ccccccc}
\hline\hline
\multirow{2}*{\minitab{PCG1}}& \multirow{2}*{\minitab{PCG2}}  & \multirow{2}*{\minitab{ ancilla qubit}}& \multicolumn{2}{c}{feedforward} \\ \cline{4-5}%
                             &                                &                           &  control  qubit   &       target  qubit\\ \hline
\multirow{4}*{\minitab{R}}   & \multirow{2}*{\minitab{L}}     &       $\uparrow$          &                   &               \\  
                             &                                &        $\downarrow$       &                   &  $\sigma_x$   \\ 
                             & \multirow{2}*{\minitab{R}}     &         $\uparrow$        &       $-\sigma_z$ &               \\ 
                             &                                &         $\downarrow$      &       $\sigma_z$  &  $\sigma_x$     \\  \hline
\multirow{4}*{\minitab{L}}   & \multirow{2}*{\minitab{L}}     &         $\uparrow$        &                   &    $\sigma_x$      \\ 
                             &                                &        $\downarrow$       &                   &             \\  
                             & \multirow{2}*{\minitab{R}}     &         $\uparrow$        &       $-\sigma_z$ &     $\sigma_x$     \\ 
                             &                                &      $\downarrow$         &       $\sigma_z$  &      \\
                             \hline\hline
\end{tabular}\label{Table}
\end{table}

BSA is an important prerequisite for many quantum protocols, such as
superdense coding, teleportation,  entanglement swapping, and so on.
Next, based on spin-QD-double-side-cavity units, we show the
principle of our complete BSA for Fermionic two-qubit systems. It
can be implemented with the left two parts shown in Fig.\ref{Fig2}.
Considering the system composed of the two excess electrons in
cavities 1 and 2. It is prepared in the four Bell states
\begin{eqnarray}  \label{eq.4}
|\psi^{\pm}\rangle = \frac{1}{\surd2}(|\uparrow_1\uparrow_2\rangle\pm |\downarrow_1\downarrow_2\rangle), \nonumber\\
|\varphi^{\pm}\rangle
=\frac{1}{\surd2}(|\uparrow_1\downarrow_2\rangle\pm
|\downarrow_1\uparrow_2\rangle).
\end{eqnarray}
After the PCG operation is performed on the qubits 1 and 2 (the
inject probe photon, labeled as $in_1$, is in state $|R_1\rangle$),
the four Bell states are divided into two groups. That is,
$|\psi^{\pm}\rangle$ and $|\varphi^{\pm}\rangle$.
$|\psi^{\pm}\rangle$ correspond  to the click of the detector $D_2$,
and $|\varphi^{\pm}\rangle$ correspond to $D_1$. The $"+" $state and
the $"-"$ state in each group can be distinguished by the same
operation of PCG on the system in the second time after a Hadamard
operation is performed on both the control and the target qubits. In
detail, the $"+"$ state corresponds to the click of the detector
$D_2$, and the $"-"$ state corresponds to $D_1$. This scheme can be
extended to create remote multi-spin entangled states such as
Greenberger-Horne-Zeilinger states (GHZ) or cluster states
\cite{multi-spin}.

By far, we have shown the principles for PCG, CNOT gates, and BSA
under the ideal condition. We consider imperfections due to side
leakage of cavity field, the trion dephasing, and the heavy-light
hole mixing.

The fidelity of the CNOT gate associates with the reflection and
transmission operators of the system.  The two operators, include
the contributions both from the uncoupled and from the coupled
cavities, can be described as \cite{Hu2}
\begin{eqnarray}    \label{eq.5}
\widehat{t}(\omega)&=&t_0(\omega)(|R\rangle\langle R|\otimes |\uparrow\rangle\langle\uparrow|
+|L\rangle\langle L|\otimes |\downarrow\rangle\langle\downarrow|)+\nonumber\\
         &&t(\omega)(|R\rangle\langle R|\otimes|\downarrow\rangle\langle\downarrow|
         +|L\rangle\langle
         L|\otimes|\uparrow\rangle\langle\uparrow|),\nonumber\\
\widehat{r}(\omega)&=&r_0(\omega)(|R\rangle\langle R|\otimes |\uparrow\rangle\langle\uparrow|
+|L\rangle\langle L|\otimes |\downarrow\rangle\langle\downarrow|)+\nonumber\\
         &&r(\omega)(|R\rangle\langle R|\otimes|\downarrow\rangle\langle\downarrow|
         +|L\rangle\langle  L|\otimes|\uparrow\rangle\langle\uparrow|).
\end{eqnarray}
with
\begin{eqnarray}    \label{eq.6}
r(\omega)&=&1+t(\omega),\nonumber\\
t(\omega)&=&\frac{-\kappa[i(\omega_{X^{-}}-\omega)+\frac{\gamma}{2}]}{[i(\omega_{X^{-}}
-\omega)
+\frac{\gamma}{2}][i(\omega_c-\omega)+\kappa+\frac{\kappa_s}{2}]+g^2}.
\end{eqnarray}
Here, $r(\omega)$ and $t(\omega)$ are the reflection and the
transmission coefficients of the coupled cavities with $g\neq0$,
respectively. $r_0(\omega)$ and $t_0(\omega)$ are the reflection and
the transmission coefficients of the uncoupled cavities with $g=0$
in Eq.(\ref{eq.6}). $\omega$, $\omega_c$, and $\omega_{X^-}$ are the
frequencies of the input photon, cavity mode, and  $X^-$ transition,
respectively. $g$ is cavity coupling strength.  $\gamma/2$,
$\kappa$, and $\kappa_s/2$ are the  $X^-$ dipole decay rate, the
cavity field decay rate, and the cavity field leaky rate,
respectively.

In our schemes, we consider the resonance with
$\omega_c=\omega_{X^-}=\omega$.  Eq.(\ref{eq.6}) can be simplified
as
\begin{eqnarray}  \label{eq.7}
t_0(\omega)=-\frac{\kappa}{\kappa+\frac{\kappa_s}{2}},\;\;\;\;
r_0(\omega)=\frac{\frac{\kappa_s}{2}}{\kappa+\frac{\kappa_s}{2}},
\end{eqnarray}
and
\begin{eqnarray}  \label{eq.8}
r(\omega)=1+t(\omega),\;\;\;\;
t(\omega)=-\frac{\frac{\gamma}{2}\kappa
}{\frac{\gamma}{2}[\kappa+\frac{\kappa_s}{2}]+g^2}.
\end{eqnarray}

For an ideal case, that is, the side leakage $\kappa_s$  is much
lower than the cavity decay rate $\kappa$, and then
 $|t_0(\omega)|\rightarrow 1$, $|r_0(\omega)|\rightarrow 0$ for
the cold cavity and $|t(\omega)|\rightarrow 0$,
$|r(\omega)|\rightarrow 1$ for the hot cavity in the strong coupling
regime $g>(\kappa,\gamma)$ \cite{Hu2}. Our scheme for a CNOT gate
can achieve a unity fidelity in the strong-coupling regime. However,
this is a big challenge for QD-micropillar cavities although
significant progress has been made \cite{challenge}.

For an unideal case, that is, the cavity side leakage $\kappa_s$,
which will cause bit-flip error, is taken into account, and then
\begin{eqnarray}    \label{eq.10}
|R^\downarrow\downarrow\rangle&\rightarrow&|r||L^\uparrow\downarrow\rangle+|t||R^\downarrow\downarrow\rangle,\nonumber\\
|L^\uparrow\downarrow\rangle&\rightarrow&|r||R^\downarrow\downarrow\rangle+|t||L^\uparrow\downarrow\rangle,\nonumber\\
|R^\downarrow\uparrow\rangle&\rightarrow&-|t_0||R^\downarrow\uparrow\rangle-|r_0||L^\uparrow\uparrow\rangle,\nonumber\\
|L^\uparrow\uparrow\rangle&\rightarrow&-|t_0||L^\uparrow\uparrow\rangle-|r_0||R^\downarrow\uparrow\rangle.
\end{eqnarray}
The fidelity of the CNOT gate can be written as
\begin{eqnarray}    \label{eq.11}
F&=&1-P,\nonumber\\
&=&\frac{(\frac{\kappa_s}{\kappa})^4+16}{(\frac{\kappa_s}{\kappa})^4+16(\frac{\kappa_s}{\kappa})^2+16}
=\frac{200(\frac{g}{\kappa})^4+1}{200(\frac{g}{\kappa})^4+3} ,
\end{eqnarray}
by taking $\gamma=0.1\kappa$ which is experimentally achieved,
$|t_0|=|r|$ (that is,
$(\frac{g}{\kappa})^2=\frac{\kappa}{10\kappa_s}-\frac{\kappa_s}{40\kappa}$)
which is required for our protocol, and
$\omega_c=\omega_{X^-}=\omega$. Here $P$ is the error rate.

\begin{figure}[!h]
\begin{center}
\includegraphics[width=6.5 cm,angle=0]{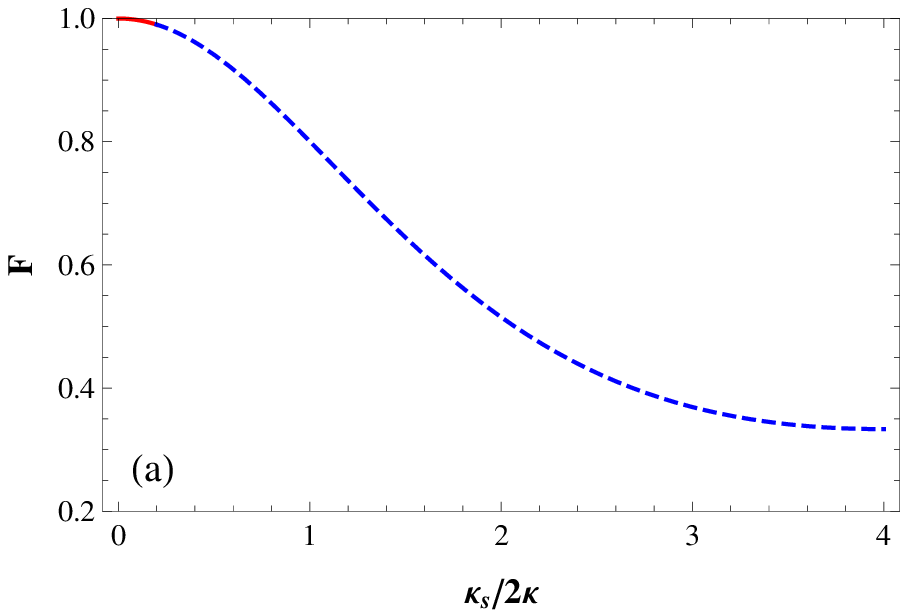}\\ \includegraphics[width=6.5 cm,angle=0]{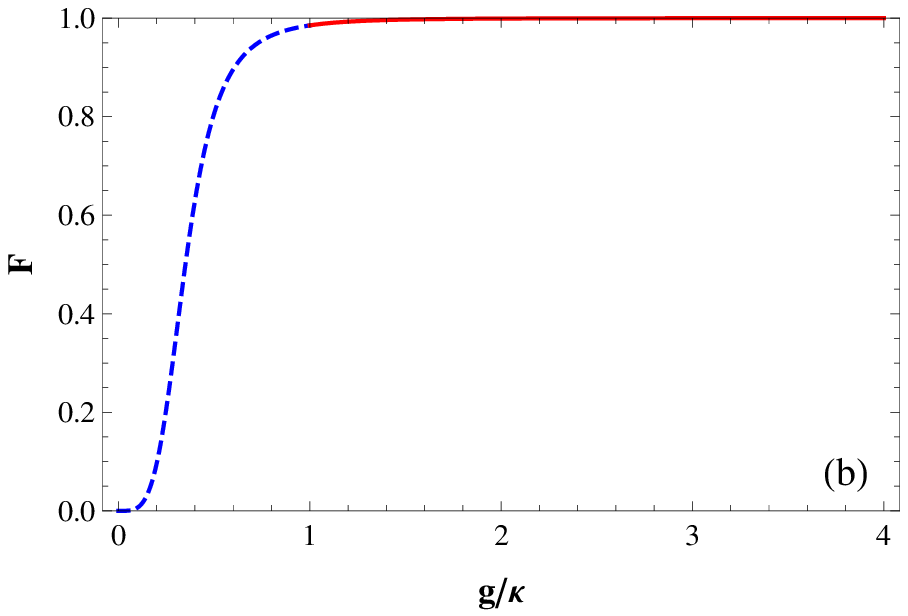}
\caption{(Color onlines) (a) The fidelity of the CNOT gate vs the
side leakage rate $\kappa_s$.  (b) The fidelity of the CNOT gate vs
the coupling strength. $\gamma=0.1\kappa$ which is experimentally
achievable and $|r|=|t_0|$ (that is,
$\frac{g^2}{\kappa^2}=\frac{\kappa}{10\kappa_s}-\frac{\kappa_s}{40\kappa}$)
which is required for our protocol are taken for (a)-(b), and
$\omega_c=\omega_{X^-}=\omega=\omega_0$ is assumed.} \label{Fig4}
\end{center}
\end{figure}

As shown in Fig.\ref{Fig4}(a) and Fig.\ref{Fig4}(b), a near-unity
fidelity of the CNOT gate can be achieved with small
$\kappa_s/2\kappa$ in the strong-coupling regime with
$g/\kappa=2.4$, which can be achieved for the In(Ga)As QD-cavity
system \cite{challenge,achieved}. The lower $\kappa_s/2\kappa$, the
higher $F$. A higher fidelity could also be achieved by taking a
lower  $\kappa_s/2\kappa$ in the weak-coupling regime
$g<(\gamma,\kappa)$.

Besides the side leakage, the  exciton dephasing  and the imperfect
optical selection rule can also reduce the fidelity \cite{Hu2}. For
exciton dephasing (the optical dephasing and the spin dephasing of
$X^-$), it reduces the fidelity by a factor \cite{Hu2}
$[1-\exp(-\tau/T_2)]$, where $\tau$ is the cavity photon lifetime
and $T_2$ is the exciton coherence time. Here, the optical dephasing
can reduce the fidelity by only a few percents as the optical
coherence time of exciton in self-assembled In(Ga)As QDs is ten
times long as the cavity photon lifetime. The former can be reach
several hundred picoseconds \cite{Opticl-deph}, however, the later
is around tens of picoseconds in the strong coupling regime for a
cavity with a Q-factor of $10^4-10^5$; the spin dephasing of the
$X^-$ which mainly arises from the hole-spin dephasing, can be
safely neglected, that is because spin coherence time is at least
three order of the magnitude longer than the cavity photon lifetime
\cite{spin-deph1,spin-deph2}. For the imperfect optical selection
rule which is caused by the heavy-light hole mixing in realistic QD
(due to the asymmetric in the QD shape and the strain field
distribution), since the  hole mixing could be reduced by
engineering the shape and the size of QDs or choosing different
types of QDs and in the valence band is in the order of a few
percents \cite{optic-selec} [e.g., for self-assembled In(Ga)As QDs],
the imperfect optical selection rule can reduce the fidelity by only
a few percents.

In summary, we have proposed  a device which can convert the spin
parity of two static electron-spin qubits confined in  charged QDs
inside double-sided microcavities into the out-coming photon
polarization information. Using two such parity-check measurements,
we construct a deterministic CNOT gate on electron-spin qubits,
allowing for deterministic scalable quantum computing in solid-state
systems. Subsequently, a possible application of the
spin-QD-double-side-cavity, a spin Bell-state analyzer was
discussed. Moreover, from the investigation on the fidelity of the
CNOT gate, one can find that our proposal  works in both the weak
coupling and the strong coupling regimes, but high fidelities are
achieved only when the ratio of the side leakage to the cavity loss
is low.

This work is supported by the National Natural Science Foundation of
China under Grant Nos. 10974020 and 11174039,  NCET-11-0031, and the
Fundamental Research Funds for the Central Universities.

\end{document}